\documentclass[aps,prl,twocolumn,superscriptaddress,calc]{revtex4-1}
\pdfoutput=1
\usepackage{graphicx}
\usepackage{amsfonts}
\usepackage{amsmath}
\usepackage{amssymb}

\newcommand{\ket}{\right\rangle}
\newcommand{\kt}[1]{\ensuremath{\left|#1\right\rangle}}
\newcommand{\upram}{\kt{\uparrow}}

\newcommand{\downram}{\kt{\downarrow}}

\newcommand{\upqram}{\kt{\uparrow,k_x=q+Q/2}}
\newcommand{\downqram}{\kt{\downarrow,k_x=q-Q/2}}
\newcommand{\upres}{\kt{\uparrow}_R}

\newcommand{\downres}{\kt{\downarrow}_R}

\newcommand{\BSO}{\mathbf{B}^{(\text{SO})}}
\newcommand{\mitCUAaddress}{Department of Physics, MIT-Harvard Center for Ultracold Atoms, and Research Laboratory of Electronics, MIT, Cambridge, Massachusetts 02139, USA}
\newcommand{\cam}{Cavendish Laboratory, University of Cambridge, J. J. Thomson Avenue, Cambridge CB3 0HE, United Kingdom}

\begin{document}

\title{Spin-Injection Spectroscopy of a Spin-Orbit Coupled Fermi Gas}

\author{Lawrence W. Cheuk}
\affiliation{\mitCUAaddress}
\author{Ariel T. Sommer}
\affiliation{\mitCUAaddress}
\author{Zoran Hadzibabic} 
\affiliation{\mitCUAaddress}
\affiliation{\cam}
\author{Tarik Yefsah}
\affiliation{\mitCUAaddress}
\author{Waseem S. Bakr}
\affiliation{\mitCUAaddress}
\author{Martin W. Zwierlein}
\affiliation{\mitCUAaddress}

\begin{abstract}
The coupling of the spin of electrons to their motional state lies at the heart of recently discovered topological phases of matter~\cite{konig2007quantum,hsieh2008topological,hasan2010topological}. Here we create and detect spin-orbit coupling in an atomic Fermi gas, a highly controllable form of quantum degenerate matter~\cite{inguscio2008ultracold,bloc08many}. We reveal the spin-orbit gap~\cite{quay2010obse} via spin-injection spectroscopy, which characterizes the energy-momentum dispersion and spin composition of the quantum states. For energies within the spin-orbit gap, the system acts as a spin diode. To fully inhibit transport, we open an additional spin gap, thereby creating a spin-orbit coupled lattice~\cite{jime2012peierls} whose spinful band structure we probe. In the presence of s-wave interactions, such systems should display induced p-wave pairing~\cite{willi2012synth}, topological superfluidity~\cite{zhang2008superfluid}, and Majorana edge states~\cite{sato2009nona}.
\end{abstract}
\maketitle
Spin-orbit coupling is responsible for a variety of phenomena, from the fine structure of atomic spectra to the spin Hall effect, topological edge states, and the predicted phenomenon of topological superconductivity~\cite{hasan2010topological, qi2011topo}. In electronic systems, spin-orbit coupling arises from the relativistic transformation of electric fields into magnetic fields in a moving reference frame. In the reference frame of an electron moving with wavevector \textbf{k} in an electric field, the motional magnetic field couples to the electron spin through the magnetic dipole interaction. This spin-orbit coupling phenomenon is responsible for lifting the degeneracy of spin states in the excited orbitals of atoms and solid-state materials such as zinc-blende structures~\cite{dress1955spin}. In a two-dimensional semiconductor heterostructure, the electric field can arise from structure or bulk inversion asymmetry~\cite{wink2003spin}, leading to magnetic fields of the form $\mathbf{B}^{(R)} = \alpha (-k_y,k_x,0)$ or $\mathbf{B}^{(D)} = \beta (k_y,k_x,0)$. The resulting spin-orbit coupling terms in the Hamiltonian are known as the Rashba~\cite{bychkov1984osci} and Dresselhaus~\cite{dress1955spin} contributions, respectively. Including a possible momentum-independent Zeeman field $\mathbf{B}^{(Z)}=(0,B^{(Z)}_y,B^{(Z)}_z)$, the Hamiltonian of the electron takes the form:
\begin{equation}
\mathcal{H} = \frac{\hbar^2 k^2}{2m} - \frac{g \mu_B}{\hbar} \mathbf{S} \cdot (\mathbf{B}^{(D)}+ \mathbf{B}^{(R)} + \mathbf{B}^{(Z)}),
\label{eqHSO}
\end{equation}
where $g$ is the electron $g$-factor, $\mu_B$ is the Bohr magneton and $\mathbf{S}$ is the electron spin. 

The energy-momentum dispersion and the associated spin texture of the Hamiltonian in Eq.~(\ref{eqHSO}) are shown in Figure 1A for $B^{(Z)}_y=0$ and $\alpha=\beta$. In the absence of a perpendicular Zeeman field $B^{(Z)}_z$, the spectrum consists of free particle parabolas for the two spin states that are shifted relative to each other in $k$-space owing to the spin-orbit interaction. For a finite field $B^{(Z)}_z$, a gap opens in the spectrum. This gap, known as the spin-orbit gap, has been recently observed in one-dimensional quantum wires~\cite{quay2010obse,nadj2012spectroscopy}. The two energy bands are spinful in the sense that the spin of an atom is locked to its momentum. 
Similar band structures have been used to explain the anomalous quantum Hall effect and predict a saturation of the Hall conductivity for Fermi energies in the gap region~\cite{xiao2010berry}. 

In this work, we engineer the Hamiltonian in Eq.~(\ref{eqHSO}) with equal Rashba and Dresselhaus strengths in an optically trapped, degenerate gas of fermionic lithium atoms via Raman dressing of atomic hyperfine states~\cite{liu2009effect,dalibard2011artificial}. Raman fields have previously been used to generate spin-orbit coupling and gauge fields in pioneering work on Bose-Einstein condensates~\cite{lin2011spin,lin2009synthetic,aidelsburger2011experimental}, and recently spin-orbit coupling in Fermi gases~\cite{wang2012spin}. Here, we directly measure the spin-orbit band structure of Eq.~(\ref{eqHSO}), as well as the rich band structure of a spin-orbit coupled lattice. For this, we introduce spin-injection spectroscopy, which is capable of completely characterizing the quantum states of spin-orbit coupled fermions, including the energy-momentum dispersion and the associated spin-texture. By tracing the evolution of quantum states in the Brillouin zone, this method is able to directly measure topological invariants, such as the Chern number in a two-dimensional system~\cite{zhao2011chern,hasan2010topological, qi2011topo}.
\begin{figure*}
\centering
\includegraphics[scale=1]{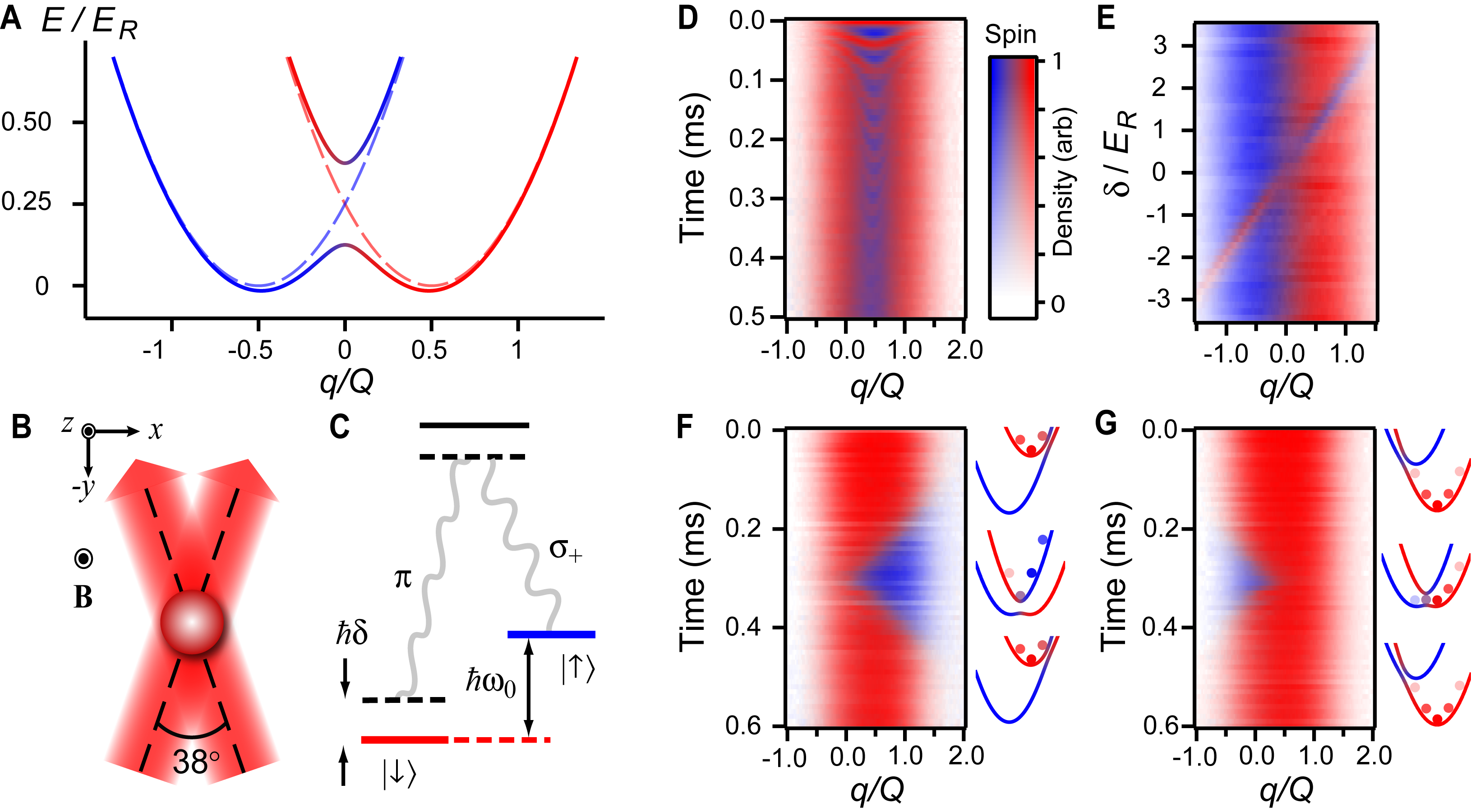}
\caption{Realization of spin-orbit coupling in an atomic Fermi gas. (A) Energy bands as a function of quasi-momentum $q$ for Raman coupling strength of $\hbar\Omega_R=0.25 E_R$ and $\hbar \delta=0$. The spin composition of the states is indicated by the color. Dashed lines show energy bands for $\hbar\Omega_R=0 E_R$ and $\hbar \delta =0 E_R$. (B) Geometry of the Raman beams: A pair of Raman beams at
$\pm 19^{\circ}$ relative to the $\hat{y}$ axis couples states $\left|\downarrow,k_x=q\right\rangle$ and $\left|\uparrow,k_x=q+Q\right\rangle$. A bias magnetic field $\mathbf B$ in the $\hat{z}$ direction provides the quantization axis. (C) The hyperfine interaction splits $\left |\uparrow\right\rangle$ and $\left|\downarrow \right\rangle$ by $\hbar \omega_0$, and the relevant polarization components are $\pi$ and $\sigma_+$. $\hbar \delta$ is the two-photon detuning. (D) Momentum-dependent Rabi oscillations in the spin texture after sudden switch-on of the Raman beams. Here $\hbar \Omega_R= 0.78(2) E_R$ and the detuning $\hbar \delta=-0.25(1)E_R$. (E) A $\pi$-pulse for the resonant momentum-class of atoms is applied at different two-photon detunings $\hbar \delta$. The Raman strength is $\hbar \Omega_R= 0.035(5) E_R$ in order to retain momentum selectivity. (F and G): Adiabatic loading and unloading of atoms into the upper (lower) band at coupling strength of $\hbar \Omega_R = 0.53(5) E_R$. The Raman beams are turned on with $\delta= \mp 8.5\Omega_R$, and the detuning is swept linearly to $\delta=0$ and back at a rate of $|\dot{\delta}|=0.27(5)\Omega_R^2$. This loads atoms into the upper (lower) band, as indicated by the diagrams on the right. The spin texture follows the instantaneous value of $\delta$, indicating adiabaticity.
}
\label{fig1f}
\end{figure*}

In order to directly reveal the single-particle eigenstates of the spin-orbit coupled system, we reduce the interactions in our Fermi gas to a negligible strength. This is convenient for studying topological insulators, whose behavior is mostly governed by single-particle physics. On the other hand, a single-component spin-orbit coupled Fermi gas is expected to develop effective p-wave interactions mediated by s-wave interactions~\cite{willi2012synth}, either in the presence of an s-wave Feshbach resonance, or in the presence of flat bands as realized below. This can lead to BCS pairing in a p-wave channel, and in a two-dimensional system with pure Rashba coupling, to $p_x+ip_y$ pairing and chiral superfluidity~\cite{willi2012synth,zhang2008superfluid}.
%

We generate spin-orbit coupling using a pair of laser beams that connect two atomic hyperfine levels, labeled $\upram$ and $\downram$, via a two-photon Raman transition (Fig.~1B,C). The Raman process imparts momentum $\hbar Q\hat{x}$ to an atom while changing its spin from $\kt{\downarrow}$ to $\kt{\uparrow}$, and momentum $-\hbar Q\hat{x}$ while changing the spin from $\kt{\uparrow}$ to $\kt{\downarrow}$. Defining a quasimomentum $q=k_x - \frac{Q}{2}$ for spin $\downram$ and $q=k_x+\frac{Q}{2}$ for spin $\upram$, one obtains the Hamiltonian of the form given in Eq.~(\ref{eqHSO})~\cite{lin2011spin}. In this mapping, $B^{(Z)}_z = \hbar \Omega_R/g \mu_B$, where $\Omega_R$ is the two-photon Rabi frequency, $B^{(Z)}_y=\hbar \delta/g \mu_B$, where $\delta$ is the two-photon detuning, and $\alpha = \beta = \frac{\hbar^2 Q}{2m g \mu_B}$ (see Supplemental Material). 

When the spin-orbit gap is opened suddenly, an atom prepared in the state $\downqram$ oscillates between $\downqram$ and $\upqram$ with a momentum dependent frequency $\Delta(q)/h$, where $\Delta(q)$ is the energy difference between the bands at quasimomentum $q$. Such Rabi oscillations correspond to Larmor precession of the pseudo-spin in the effective magnetic field $\BSO = \mathbf{B}^{(\text{D})} + \mathbf{B}^{(\text{R})} + \mathbf{B}^{(Z)}$. We have observed these oscillations by starting with atoms in $\downram$, pulsing on the Raman field for a variable duration $\tau$, and imaging the atoms spin-selectively after time-of-flight expansion from the trap. Time-of-flight maps momentum to real space, allowing direct momentum resolution of the spin populations. As a function of pulse duration, we observe oscillations of the pseudospin polarization with momentum-dependent frequencies (Fig.~1D). Our Fermi gas occupies a large range of momentum states with near-unity occupation. Therefore, each image at a given pulse duration $\tau$ contains information for a large range of momenta $q$. The observation of momentum-dependent oscillations demonstrates the presence of a spin-orbit gap, and shows that the atomic system is coherent over many cycles. To highlight the momentum selectivity of this process, we prepare an equal mixture of atoms in states $\upram$ and $\downram$ and pulse on the Raman fields for a time $t=\pi/\Omega_{R}$ for different two-photon detunings $\delta$. This inverts the spin for atoms with momentum $q$ where $\Delta(q)$ is minimal. Since the minimum of $\Delta(q)$ varies linearly with $\delta$ due to the Doppler shift $\propto k_x Q$, the momentum $q$ at which the spin is inverted depends linearly on $\delta$ (Fig.~1E).

\begin{figure*}
\centering
\includegraphics[scale=1]{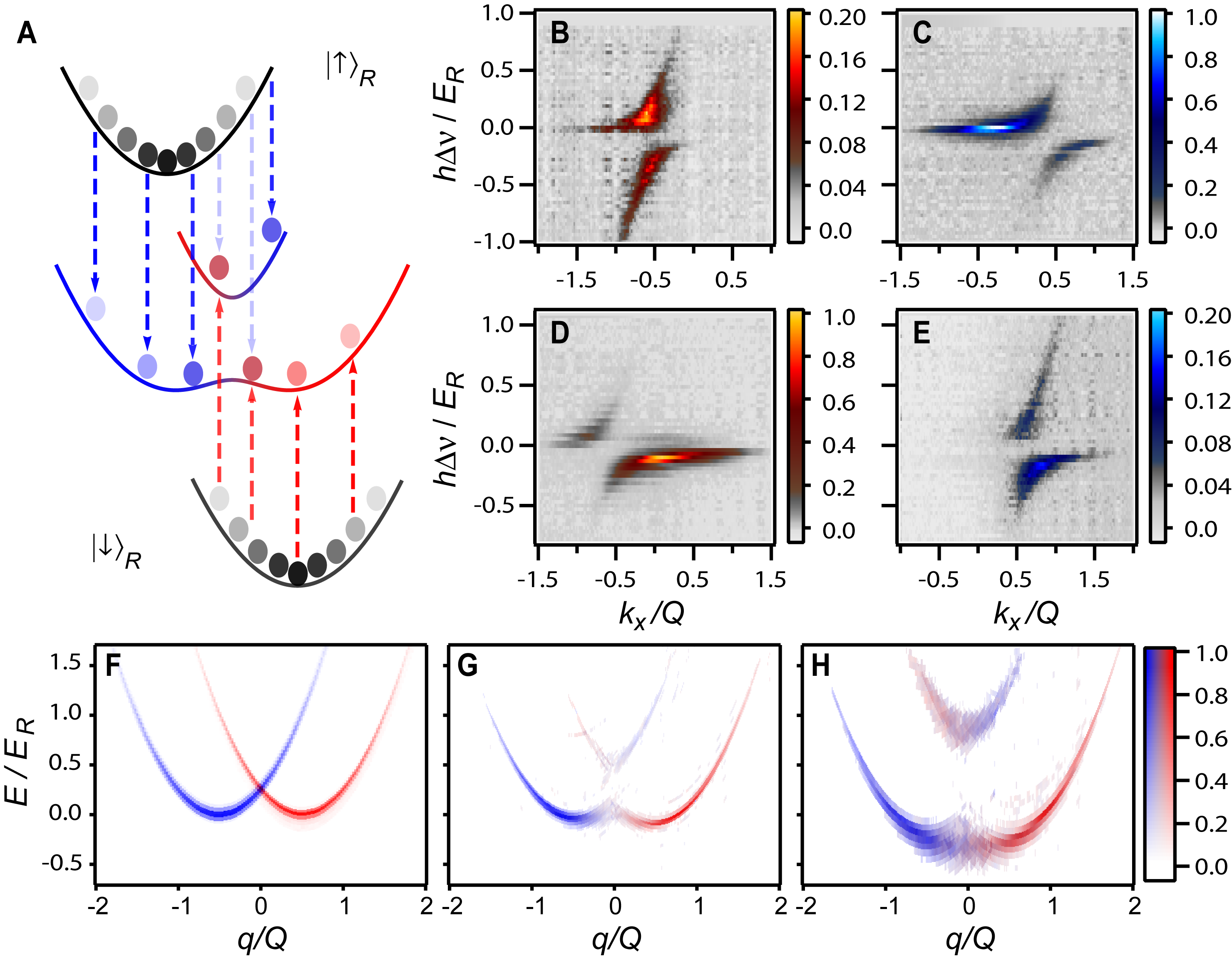}
\caption{Spin--injection spectroscopy of a spin-orbit coupled Fermi gas. (A) A radiofrequency (RF) pulse transfers atoms from the reservoir states (shown in black) $\left|\uparrow \right \rangle_R$ and $\left|\downarrow \right \rangle_R$ into the spin-orbit coupled system (shown in red and blue). Transfer occurs when the RF photon energy equals the energy difference between the reservoir state and the spin-orbit coupled state at quasi-momentum $q$. (B,C,D and E) Transfer as a function of RF frequency detuning $h\Delta \nu$ and quasi momentum $q$ at Raman coupling strength of $\hbar \Omega_R = 0.43(5) E_R$ and $\hbar \delta=0.00(3) E_R$. Note that starting with reservoir $\left|\downarrow\right\rangle_R (\left|\uparrow\right\rangle_R)$, transfer to state $\left|\uparrow\right\rangle (\left|\downarrow\right\rangle)$ is entirely due to spin-orbit coupling. Hence the signal is generally much weaker than that for state $\left|\downarrow\right\rangle  (\left|\uparrow\right\rangle)$ except right in the gap, where their ratio approaches $50\%/50\%$. (B and C) Spin-resolved $\left|\downarrow\right\rangle$ and $\left|\uparrow \right\rangle$ spectra, respectively, when transferring out of $\left|\uparrow\right\rangle_R$. (D and E) Spin-resolved $\left|\downarrow\right\rangle$ and $\left|\uparrow \right\rangle$ spectra, respectively, when transferring out of $\left|\downarrow\right\rangle_R$. (F,G and H) The reconstructed spinful dispersions for $\hbar \delta=0.00(3) E_R$ and $\hbar \Omega_R = 0 E_R$, $\hbar \Omega_R = 0.43(5) E_R$ and $\hbar \Omega_R = 0.9(1) E_R$, respectively.}
\label{fig2f}
\end{figure*}

Instead of pulsing on the Raman field and projecting the initial state into a superposition of states in the two bands, one can introduce the spin-orbit gap adiabatically with respect to band populations. This is achieved by starting with a spin-polarized Fermi gas and sweeping the two-photon detuning $\delta$ from an initial value $\delta_i$ to a final detuning $\delta_f$. The magnitude of the initial detuning $|\delta_i|$ is much larger than the two-photon recoil energy $E_R = \hbar^2  Q^2/2m$, so that the effective Zeeman field is almost entirely parallel with the spins. Depending on the direction of the sweep, this loads atoms into either the upper or the lower dressed band. We interrupt the sweep at various times, and image the spin-momentum distribution. This reveals that the spin texture follows the effective Zeeman field. The process is reversible, as we verify by sweeping the detuning back to $\delta_i$ and restoring full spin-polarization. (Fig.~1F and G).

Having demonstrated the ability to engineer spin-orbit coupling in a Fermi gas, we introduce a general approach to measure the complete eigenstates and energies of fermions at each quasi-momentum $q$ and thus resolve the band structure and associated spin texture of spin-orbit coupled atomic systems. Our approach yields equivalent information to spin and angle-resolved photoemission spectroscopy (spin-ARPES), a powerful technique recently developed in condensed matter physics~\cite{hoesch2002spinarpes}. Spin-ARPES is particularly useful for studying magnetic and quantum spin Hall materials; it has been used, for example, to directly measure topological quantum numbers in the Bi$_{1-x}$Sb$_x$ series, revealing the presence of topological order and chiral properties~\cite{hsieh2009observation}. 

Our spectroscopic technique uses radiofrequency (RF) spin-injection of atoms from a free Fermi gas into an empty spin-orbit coupled system using photons of a known energy (Fig.~2A). After injection, the momentum and spin of the injected atoms are analyzed using time of flight~\cite{stewart2008using} combined with spin-resolved detection. Atoms are initially loaded into one of two free ``reservoir" atomic states $\downres$ and $\upres$ which can be coupled to the states $\downram$ and $\upram$, respectively, via the RF spin-injection field, without changing the quasimomentum. The injection occurs when the frequency of the RF pulse matches the energy difference between the spin-orbit coupled bands and the initial reservoir state (see Fig.~2A). Spin-injection from $\downres$ ($\upres$) populates mostly the region of the spin-orbit coupled bands with a strong admixture of $\downram$ ($\upram$) states. Thus, the use of two reservoir states allows us to measure both the $\downram$-rich and the $\upram$-rich parts of the spin-orbit coupled bands. Following the injection process, the Raman beams are switched off, and the atoms are simultaneously released from the trap. After a sufficiently long time of free expansion, the density distribution gives access to the momentum distribution, which we measure using state-selective absorption imaging. By counting the number of atoms of a given spin and momentum as a function of injection energy, we determine the dispersion of the spin-orbit coupled bands along with their spin texture. 

\begin{figure*}
\centering
\includegraphics[scale=1]{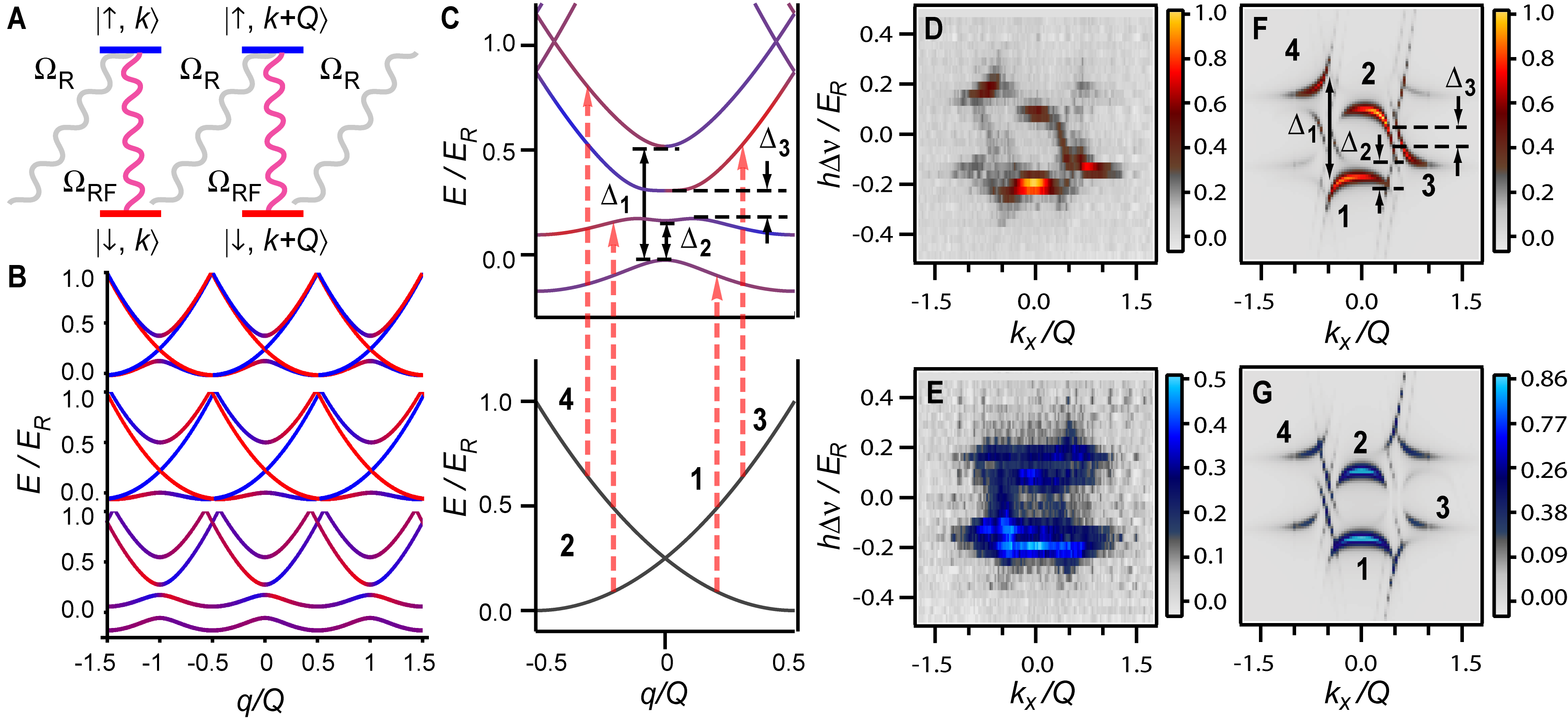}
\caption{Creating and probing a spin-orbit coupled lattice. (A) The addition of a radiofrequency field allows momentum transfer of any multiple of $Q$. The combined Raman-RF system produces a spinful lattice band structure. (B) The band structure of the Raman-RF system in the repeated zone scheme. The topmost band structure corresponds to $\hbar \Omega_{RF}=0$ and $\hbar \Omega_R=0.25E_R$, which has a band crossing at quasi-momentum $q=0$. The middle band structure corresponds to a larger Raman coupling of $\hbar \Omega_{R}=0.5 E_R$ with $\hbar \Omega_{RF}=0$. In the bottom-most band structure, $\hbar \Omega_{R}=0.5E_R$ while $\hbar \Omega_{RF}$ is increased to $0.25E_R$. (C) Spin injection from free particle bands to spinful lattice bands, starting from $\left|\downarrow \right\rangle_R$. Transitions near zero RF detuning ($h\Delta \nu \sim 0$) that give rise to dominant spectral features are identified. (D and E) Experimental spectrum of the Raman-RF system with $\hbar\Omega_R=0.40(5) E_R$ and $\hbar\Omega_{RF}=0.28(2) E_R$ in the spin $\downram$ and spin $\upram$ channels, measured after injection from reservoir $\left|\downarrow\right\rangle_R$. The dominant features span many Brillouin zones, corresponding to projection of lattice states onto free particle states after time-of-flight. (F and G)  The theoretical spectra corresponding to D and E, respectively. The features corresponding to the gaps and transitions identified in C are labeled.}
\label{fig3f}
\end{figure*}

The topological characteristics of the bands, which are encoded in the eigenstates, can be extracted from the spin and momentum composition. For our spin-orbit system with $\delta=0$, the spin of the eigenstates is confined to the $y$-$z$ plane on the Bloch sphere because the effective magnetic field has no $\hat{x}$ component. More general couplings may not restrict the spin to a great circle on the Bloch sphere, in which case at least two spin components must be measured for a complete characterization of the bands. This can be achieved by rotating the different spin components onto the measurement basis with an RF pulse.

Applying spin-injection spectroscopy, we have measured the band structure of the equal-part Rashba-Dresselhaus Hamiltonian at $\delta=0$ for several $\Omega_R$. Figure~2B, C, D and E show spin- and momentum- resolved spin-injection spectra obtained with atoms starting in the $\upres$ reservoir (top row) and starting in the $\downres$ reservoir (bottom row), for the case $\hbar\Omega_R=0.43(5)E_R$ and $\delta=0$. The $(q,\uparrow) \leftrightarrow (-q,\downarrow)$ symmetry of the system can be seen in the spectra in Fig.~2. The energy at each quasimomentum is found by adding the energy injected into the system by the RF pulse to the initial kinetic energy of the free particle in the reservoir. Figure~2F, G and H show the dispersion and spin texture of the bands obtained from the data. As $\Omega_R$ is increased, we observe the opening of a spin-orbit gap at $q=0$. The spin composition of the bands evolves from purely $\kt{\uparrow}$ or $\kt{\downarrow}$ away from the spin-orbit gap to a mixture of the two spin states in the vicinity of the spin-orbit gap, where the spin states are resonantly coupled.

The dispersion investigated above is the simplest possible for a spin-orbit coupled system and arises naturally in some condensed matter systems. A Fermi gas with this dispersion has an interesting spinful semi-metallic behavior when the Fermi energy lies within the spin-orbit gap. When the Fermi energy is outside the spin-orbit gap, there is a four-fold degeneracy of states at the Fermi surface. Inside the gap, however, the degeneracy is halved. Furthermore, propagation of spin up particles at the Fermi energy can only occur in the positive $q$ direction, while spin down fermions can only propagate in the opposite way. Particles are thus protected from back-scattering in the absence of magnetic impurities that would rotate their spin. Such a spinful semi-metal can be used to build spin-current diodes, since the material permits flow of polarized spin-currents in one direction only.

An even richer band structure involving multiple spinful bands separated by fully insulating gaps can arise in the presence of a periodic lattice potential. This has been realized for Bose-Einstein condensates by adding RF coupling between the Raman-coupled states $\left|\uparrow\ket$ and $\left|\downarrow\ket$~\cite{jime2012peierls}. Using a similar method, we create a spinful lattice for ultracold fermions, and use spin-injection spectroscopy to probe the resulting spinful band structure. The combined Raman/RF coupling scheme is shown in Fig.~3A. The Raman field couples the states $\left|\downarrow,k_x=q\ket$ and $\left|\uparrow,k_x=q+Q\ket$ with strength $\Omega_R$, whereas the RF field couples the states $\left|\downarrow,k_x=q\ket$ and $\left|\uparrow,k_x=q\ket$ with strength $\Omega_{RF}$. As a result, the set of coupled states for a given quasimomentum $q$, shown in the repeated Brillouin scheme in Fig.~3B, is $\left|\sigma, k_x=q+nQ\ket$ for integer $n$ and $\sigma=\uparrow,\downarrow$. The lowest four bands are degenerate at the band center $q=0$ when $\Omega_R=\Omega_{RF}=0$. The Raman field splits the degeneracy between the first and fourth band, leaving the other two degenerate. The remaining degeneracy, which is a Dirac point, is removed with the addition of the RF field. Thus, when the system is filled up to the top of the second band, it is an insulator. Furthermore, when $\Omega_{RF}$ is large enough, a band gap also opens between the first and second bands. 

Fig.~3D and E show the spin-injection spectra, measured with fermions initially in reservoir state $\downres$, which is sufficient to reconstruct the full band structure given the $(q,\uparrow) \leftrightarrow (-q,\downarrow)$ symmetry of the Hamiltonian. The transitions between the reservoir and the spin-orbit coupled bands for $\hbar\Omega_R=0.40(5) E_R$ and $\hbar\Omega_{RF}=0.28(2) E_R$ are shown in Fig.~3C. The experimental spectra (Fig.~3D and E) for the same parameters are compared to the corresponding theoretically calculated spectra, shown in Fig.~3F and G. We focus on the features of the $\downram$ channel of the spectrum, which is stronger because of the better spin-composition overlap with the reservoir state. 
The spectrum exhibits four prominent features separated by three energy gaps, labeled $\Delta_1$, $\Delta_2$ and $\Delta_3$ in Fig.~3F and 3G. The gaps giving rise to these features are shown on the band structure in Fig.~3C. The gap $\Delta_1$ is opened by the spin-orbit coupling, while $\Delta_2$ is opened by a direct RF coupling and $\Delta_3$ is opened by a second order process that involves both the RF and Raman fields, explaining its smallness. We have explored the Raman/RF system for a range of coupling strengths as shown in the spectra in Fig.~4B and 4C. The corresponding band structures are shown in Fig~4A. With a careful choice of the Raman/RF coupling strengths, spinful flat bands are realized, where interactions should play a dominant role~\cite{kai2011nearly}.

\begin{figure*}[htb]
\centering
\includegraphics[scale=1]{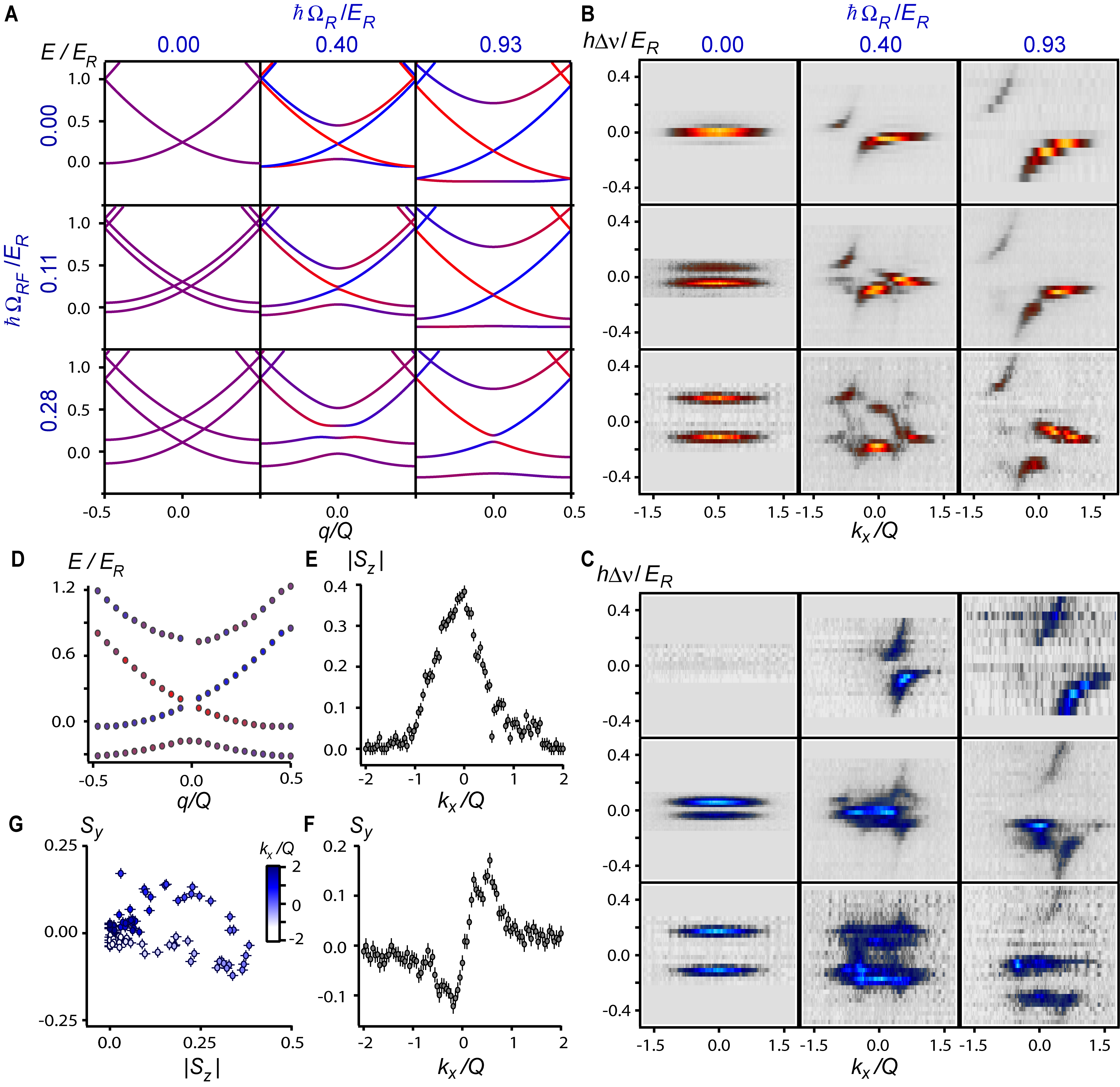}
\caption{Evolution of spin-textured energy bands of a spin-orbit coupled lattice. (A) Theoretical band structures for various combinations of $\Omega_R$ and $\Omega_{RF}$. The first band becomes flat while remaining spinful for $\hbar\Omega_R=0.93E_R$ and $\hbar\Omega_{RF}=0.11E_R$ and $0.28E_R$. (B and C) The corresponding experimental Raman-RF spin-injection spectra for injection from $\left|\downarrow\right\rangle_R$ for channels $\left|\downarrow\right\rangle$ and $\left|\uparrow \right\rangle$, respectively. The color map used is the same as Fig.~2B and 2E after rescaling to the maximum intensity (which for 4C is 20$\%$ of 4E), except for the top left panel in (C), which is scaled to the maximum intensity of the corresponding panel in (B). Possible interaction effects between $\left|\uparrow\right\rangle$ with $\left|\downarrow\right\rangle_R$ (see Fig. S2) makes only the dominant features resolvable in $\left|\uparrow\right\rangle$, while finer features are visible in $\left|\downarrow\right\rangle$. (D) Experimentally reconstructed band structure for $\hbar \Omega_R=0.93(7) E_R$ and $\hbar \Omega_{RF}=0.28(2)$. The spin texture is indicated by the color of the points. (E,F and G) Experimentally measured spin components $S_y$ and $|S_z|$ as a function of momentum $k_x$ for the lattice wavefunctions corresponding to the bottommost band in D.}
\label{fig4f}
\end{figure*}

To illustrate how the energy bands along with the corresponding eigenstates can be extracted, we reconstruct the energy bands along with the spin texture for $\hbar\Omega_R=0.93(7) E_R$ and $\hbar\Omega_{RF}=0.28(2) E_R$, as shown in Fig.~4D. The energies of the bands are obtained from the resonant frequencies in the spin-injection spectra, while the spin composition is extracted from the relative weights of the signal in the two spin channels (see Supplemental Material). In general, the eigenvector $\left|\psi^{(n)}(q)\right\rangle$ for the $n$th energy band at a given quasi-momentum $q$, can be expanded in terms of free space eigenstates as $\left|\psi_n(q)\right\rangle = \sum_{m,\sigma} c^{(n)}_{\sigma}(k_x = q+mQ)\left|\sigma, k_x = q+mQ\right\rangle$. In spin-injection spectroscopy, the projection of the lattice wavefunctions onto free particle states allows us to not only extract the average spin, but also the magnitude of the coefficients $c^{(n)}_{\sigma}(k_x)$. From the projection coefficients $c^{(n)}_{\sigma}(k_x)$, one can define the spin $\vec{S}(k_x)$ (see Supplemental materials). In Fig.~4E, F and G, we show the extracted value of $S_y(k_x)$ and $|S_z(k_x)|$ for the bottommost band when $\hbar\Omega_R=0.93(7) E_R$ and $\hbar\Omega_{RF}=0.28(2) E_R$. For more general spin-orbit Hamiltonians involving $\sigma_x$, one can extract the phase between all three components of $\vec{S}(k_x)$ with additional RF pulses, and fully characterize the eigenstate for the corresponding quasimomentum $q$. The topology of the band, encoded in the evolution of its eigenstates across the Brillouin zone, can thus be measured.

In summary, we have created and directly probed a spin-orbit gap in a Fermi gas of ultracold atoms and realized a fully gapped band structure allowing for spinful flat bands. We introduced spin-injection spectroscopy to characterize the spin-textured energy-momentum dispersion. Such measurements would reveal the non-trivial topology of the bands in systems with more general spin-orbit couplings~\cite{sau2011chiral}, opening a path to probing topological insulators with ultracold atoms. Recently developed high numerical aperture imaging techniques can be used for microscopic patterning of lower dimensional Fermi gases into heterostructures with regions characterized by different topological numbers separated by sharp interfaces~\cite{bakr2010probing, sherson2010single}. In such systems, spatially resolved spin-injection spectroscopy can directly reveal topologically protected edge states such as Majorana fermions, which have been proposed for topological quantum computation\cite{tewa2007quan,sato2009nona,jiang2011majorana}.

\begin{acknowledgments}
This work was supported by the NSF, a grant from the Army Research Office with funding from the DARPA OLE program, ARO-MURI on Atomtronics, AFOSR-MURI, ONR YIP, DARPA YFA, an AFOSR PECASE, and the David and Lucile Packard Foundation. Z. H. acknowledges funding from EPSRC under Grant No. EP/I010580/1.
\end{acknowledgments}
\clearpage
\section{Supplemental Materials}

\subsection{System preparation and Raman setup}
Fermionic $^6$Li in the hyperfine state $\kt{F=3/2,m_F=3/2}$ is sympathetically cooled to degeneracy by $^{23}$Na atoms in a magnetic trap. The atoms are transferred to a nearly spherical crossed optical dipole trap with mean trapping frequency $\sim 150$~Hz. Depending on the measurement, the atoms are then transferred into one of the lowest four hyperfine states in the ground state manifold via radiofrequency (RF) sweeps. The four lowest hyperfine states are states $\kt{\downarrow}_R$, $\kt{\downarrow}$, $\kt{\uparrow}$, $\kt{\uparrow}_R$ in the text. Subsequently, the magnetic field is ramped to $\mathbf{B}=B_0 \hat{z}$, with $B_0=11.6$~G. 
The geometry of the Raman beams is shown in Fig.~1B. The two beams, detuned by 3.96~GHz to the blue of the D$_1$ line, generate a moving lattice with lattice wavevector $Q = 2\pi \times (1.0 \mu\text{m})^{-1}$ and corresponding recoil energy of $E_R = \frac{\hbar^2 Q^2}{2m} = h \times 32(1)$~kHz. The calibration of the recoil energy $E_R$ is performed using the data in Fig.~1E and relies only on the relation of $Q$ to the traveled distance for a given time-of-flight. To couple states $\kt{\uparrow}$ and $\kt{\downarrow}$, the frequency difference between the two beams is set near the hyperfine splitting of $\omega_0$ = $2\pi \times $ 207.7~MHz. For the one-photon detuning that we use, $\Omega_R/\Gamma_{sc} \approx 240$, where $\Gamma_{sc}$ is the scattering rate. Note that one-body losses from single-photon scattering events do not perturb the measured spin-injection spectra, as they affect all energy states equally and only reduce the number of atoms in each momentum state. 

\subsection{Experimental procedure}
In the presence of the Raman beams, a differential Stark shift between $\kt{\uparrow}$ and $\kt{\downarrow}$ can alter the resonant two-photon frequency. The frequency corresponding to $\delta=0$ is calibrated using RF spectroscopy on the $\kt{\uparrow}$ to $\kt{\downarrow}$ transition in the presence of the Raman beams with a large two-photon detuning $\delta \approx 2\pi \times $1~MHz. The resonant frequencies for $\kt{\downarrow}_R\rightarrow \kt{\downarrow}$ and $\kt{\uparrow}_R\rightarrow \kt{\uparrow}$ are calibrated similarly. The differential Stark shift for the largest Raman coupling strength is measured to be 4~kHz for $\kt{\uparrow} \rightarrow \kt{\downarrow}$, and $<1$~kHz for the $\kt{\downarrow}_R\rightarrow \kt{\downarrow}$ and $\kt{\uparrow}_R\rightarrow \kt{\uparrow}$ transitions. For spin-injection spectroscopy, the injection process uses a RF field that couples the states $\kt{\downarrow }_R (\kt{\uparrow }_R)$ and $\kt{\uparrow }(\kt{\downarrow }_R)$. All spectra are taken with a RF injection pulse duration of 0.5~ms, and an injection field strength corresponding to a maximum transfer fraction $< 0.30$. The experimental frequency resolution for spin-injection is $ 3$~kHz $\approx 0.1E_R$, while the momentum resolution is estimated to be $ 0.05 Q$, limited by expansion time and imaging resolution. For all measurements, state-selective absorption images are taken after time-of-flight of 4~ms. As the atoms are released, the bias magnetic field is ramped to $ 300$~G, where the resonant imaging frequencies for different hyperfine states are well-resolved, allowing spin-selective absorption imaging. To obtain the spectra shown in Fig.~1 and 2, the time-of-flight images for each spin state are first integrated along $\hat{y}$, orthogonal to the spin-orbit direction. For a given quasi-momentum $q$, the integrated one-dimensional density profiles from the two spin channels are then combined to produce the final spectrum. As an example, we show in Fig.~S1 the time-of-flight images and the corresponding integrated density profiles, at a specific detuning $\delta/E_R$ for the spectrum in Fig.~1E. 

\setcounter{figure}{0}
\renewcommand{\thefigure}{S\arabic{figure}}
\begin{figure*}
\centering
\includegraphics[scale=1]{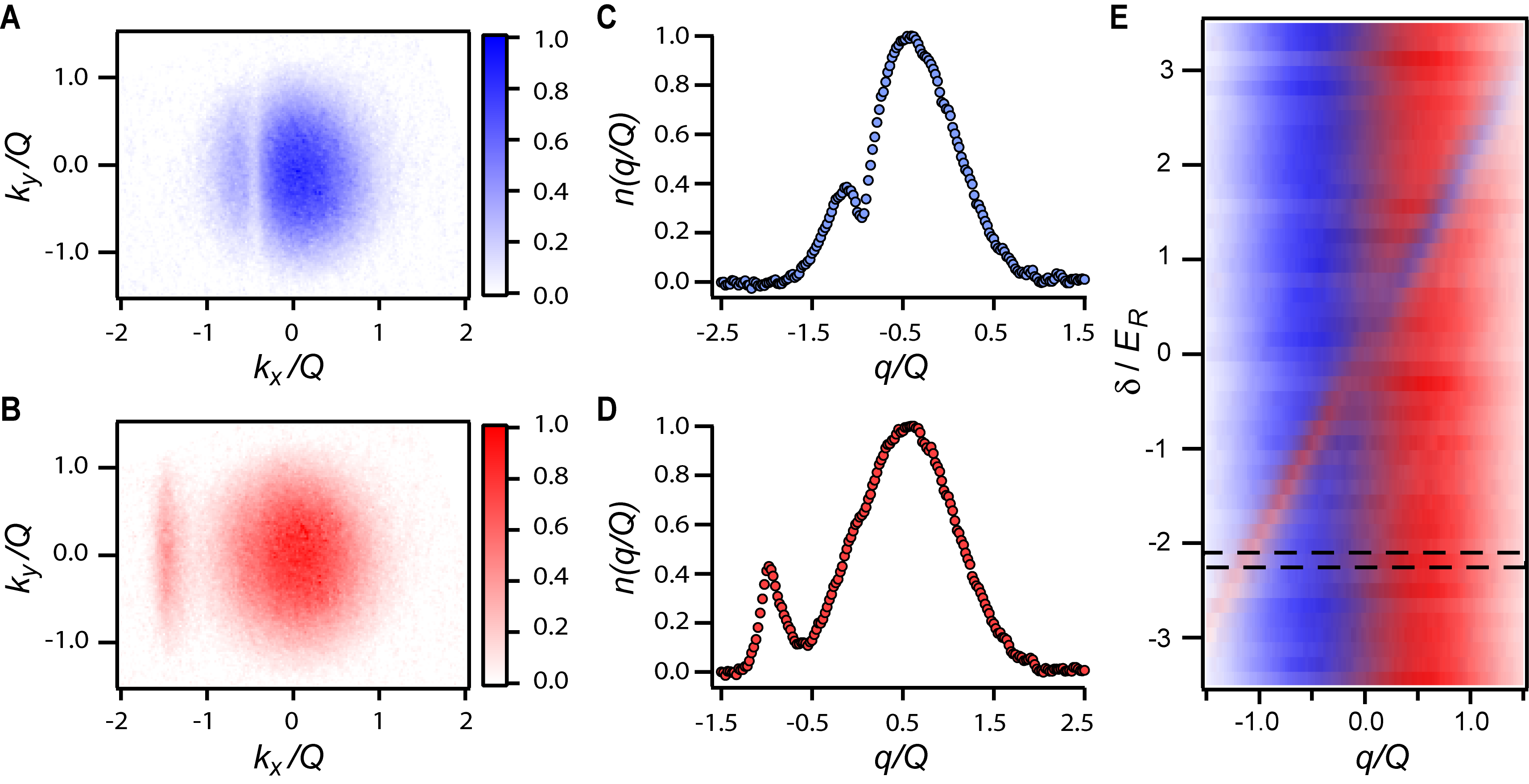}
\caption{Example of converting time-of-flight images to a spinful spectrum. Here, we show the conversion process for the spectrum in Fig.~1E for a specific detuning $\hbar \delta = -1.88 E_R$. (A and B) Time-of-flight images rescaled in terms of the recoil momentum $Q$ in the $\left|\uparrow\right\rangle$ and $\left|\downarrow\right\rangle$ channels, respectively. (C and D) The time-of-flight images in A and B are integrated along the $k_y$ direction to produce 1-dimensional densities $n(q/Q)$. The conversion to quasi-momentum $q$ involves adding a spin channel dependent momentum offset. (E) Combining C and D for every quasi-momentum $q$ produces a slice in the final spectrum. In the figure, this is the region bounded by the two dashed lines. Repeating the same procedure for other detunings yields the full spectrum.}
\label{figS2f}
\end{figure*}

\subsection{Hamiltonian for Raman-coupled system}
The spin of an atom is coupled to its momentum using a pair of laser beams near a Raman transition. Using the rotating wave approximation, the Raman beams generate a spinor potential
\begin{equation}
V(\vec{r}) = \frac{\hbar \Omega_R}{2} (\sigma_x \cos Qx - \sigma_y \sin Qx)+\frac{\delta}{2} \sigma_z,
\end{equation}
where $\Omega_R$ is the two-photon Rabi frequency.
After a local pseudo-spin rotation about the $z$ axis with angle $Qx$, the Hamiltonian becomes~\cite{lin2011spin}
\begin{equation}
\mathcal{H} = \frac{\hbar^2 \mathbf k^2}{2 m} +\frac{\hbar^2 Q}{2m}\sigma_z q  +\frac{\hbar \Omega_R}{2} \sigma_x+\frac{\delta}{2} \sigma_z + \frac{E_R}{4},
\end{equation}
where $q$ is the quasi-momentum defined in the text.
Following a global pseudo-spin rotation $\sigma_z\rightarrow \sigma_y$, $\sigma_y \rightarrow \sigma_x$ and $\sigma_x \rightarrow \sigma_z$, the Hamiltonian becomes
\begin{equation}
\mathcal{H} = \frac{\hbar^2 \mathbf k^2}{2 m} +\frac{\hbar^2 Q}{2m}\sigma_y q  +\frac{\hbar \Omega_R}{2} \sigma_z+\frac{\delta}{2} \sigma_y + \frac{E_R}{4},
\end{equation}
which up to a constant has the same Rashba-Dresselhaus form as Eq.~(1), with $\alpha=\beta=\frac{\hbar^2 Q}{2mg \mu_B}$, $B_y^{(Z)}=\hbar \delta/g \mu_B$ and $B_z^{(Z)} = \hbar \Omega_R/g \mu_B$. In this convention, the bare hyperfine states, labeled $\left|\uparrow\right\rangle$ and $\left|\downarrow\right\rangle$ are eigenstates of $\sigma_y$.

\subsection{Reconstructing the spinful dispersion for the Raman-coupled system}
For the equal Rashba-Dresselhaus system, the two channels in a spectrum from reservoir $\left|\sigma\right\rangle_R$ are first relabeled by quasi-momentum $q$. The ratio of the transferred atoms in each channel at a given $q$ directly measures the $q$-dependent spin composition. The dispersion is then reconstructed by adding the free particle dispersion $\epsilon^0(q) = (q\mp Q/2)^2/2m$ to the spectrum corresponding to injection from $\left|\uparrow\right\rangle_R (\left|\downarrow \right\rangle_R)$. The result is shown in Fig.~2F,G and H, where the color denotes the spin texture and the strength of the color is weighed by the total number of atoms at a given $q$.

\subsection{Hamiltonian for the Raman/RF system}
Raman dressing creates a spin-orbit gap in momentum space. Adding radiofrequency (RF) coupling with zero momentum transfer creates a lattice potential with true band gaps \cite{jime2012peierls}. The RF drive is applied at the same frequency as the Raman frequency, with coupling strength $\Omega_{RF}$. An atom can now receive an arbitrary number of units of the recoil momentum $\hbar Q$ by interacting alternately with the Raman field and the RF field (see Figure 3(A)). When the two spin states are coupled at different momenta using Raman lasers, and additionally at the same momentum using RF, the Hamiltonian becomes
\begin{equation}
\mathcal{H} = \frac{\hbar^2 k^2}{2 m} + \frac{\hbar \Omega_R}{2} (\sigma_x \cos Qx - \sigma_y \sin Qx)+ \frac{\hbar \Omega_{RF}}{2} \sigma_x  + \frac{\delta}{2} \sigma_z.
\label{RamanRFH}
\end{equation}
Since the Hamiltonian has discrete translational symmetry along $\hat{x}$, its eigenstates can be expanded in plane waves as
\begin{equation}
\left|\psi_n(k_x) \right\rangle = \sum_{j,\sigma = \uparrow,\downarrow} c_\sigma^{(n)}(k_x) \left|\sigma, k_x = \tilde{k}_x + j Q \right \rangle,
\end{equation}
where $\tilde{k}_x$ is the quasi-momentum given by to $k_x$ restricted to the first Brillouin zone, and $n$ is the band index.

\subsection{Spin-injection spectrum for the Raman/RF system}
Starting from reservoir $\sigma= \uparrow, \downarrow$, the injected population in band $n$ with quasi-momentum $\tilde{k}_x$ at a given RF frequency $\omega$, $P_{I,\sigma}(\omega, n,m, \tilde{k}_x)$, is given by
\begin{eqnarray}
P_{I,\sigma}(\omega, n ,l, \tilde{k}_x) &\propto& \Omega_{I,\sigma}^2 \sum_{j} n_{\sigma} (\tilde{k}_x + l Q) |c_{\sigma}^{(n)}(\tilde{k}_x + j Q)|^2 \nonumber \\
&&\times \mathcal{L}((\hbar \omega + \epsilon^0_l(\tilde{k}_x+j Q)) - (\hbar\omega_0+ \epsilon_n(\tilde{k}_x)) ),
\end{eqnarray}
where $\Omega_{I,\sigma}$ is the RF strength coupling the reservoir state to $\left|\sigma\right\rangle$, $n_{\sigma}(k)$ is the trap-averaged momentum distribution for reservoir state $\left|\sigma\right\rangle_R$, $\epsilon^0_l(k) = \frac{\hbar^2 (k+l Q)^2}{2m}$ is the free particle dispersion of the $l$th non-interacting band, $\epsilon_n(k)$ is the dispersion for the $n$th band, $\omega_0$ is the hyperfine frequency difference between $\left|\uparrow\right \rangle$ and $\left|\downarrow \right\rangle$, and $\mathcal{L}(x)$ is the RF lineshape.
After injection, the atoms are released from the trap. After sufficient time-of-flight, the momentum distribution is given by the real space atomic density profile, which for the spin $\sigma'$ channel is 
\begin{equation}
P_{\text{TOF},\sigma',\sigma}(\omega,k_x) = \sum_{n,l} P_{I,\sigma}(\omega,n,l,\tilde{k}_x) |c^{(n)}_{\sigma'}(k_x)|^2.  
\label{RamanRFspectra}
\end{equation}
The theoretical spectra in Fig.~3F and G are obtained using Eq.~(\ref{RamanRFspectra}) and coefficients  $c^{(n)}_{\sigma}(k_x)$ found by numerically diagonalizing the Hamiltonian in Eq.~(\ref{RamanRFH}). The experimental and theoretical spectra for Fig.~4 are shown in Fig.~S2.

\begin{figure*}
\centering
\includegraphics[scale=1]{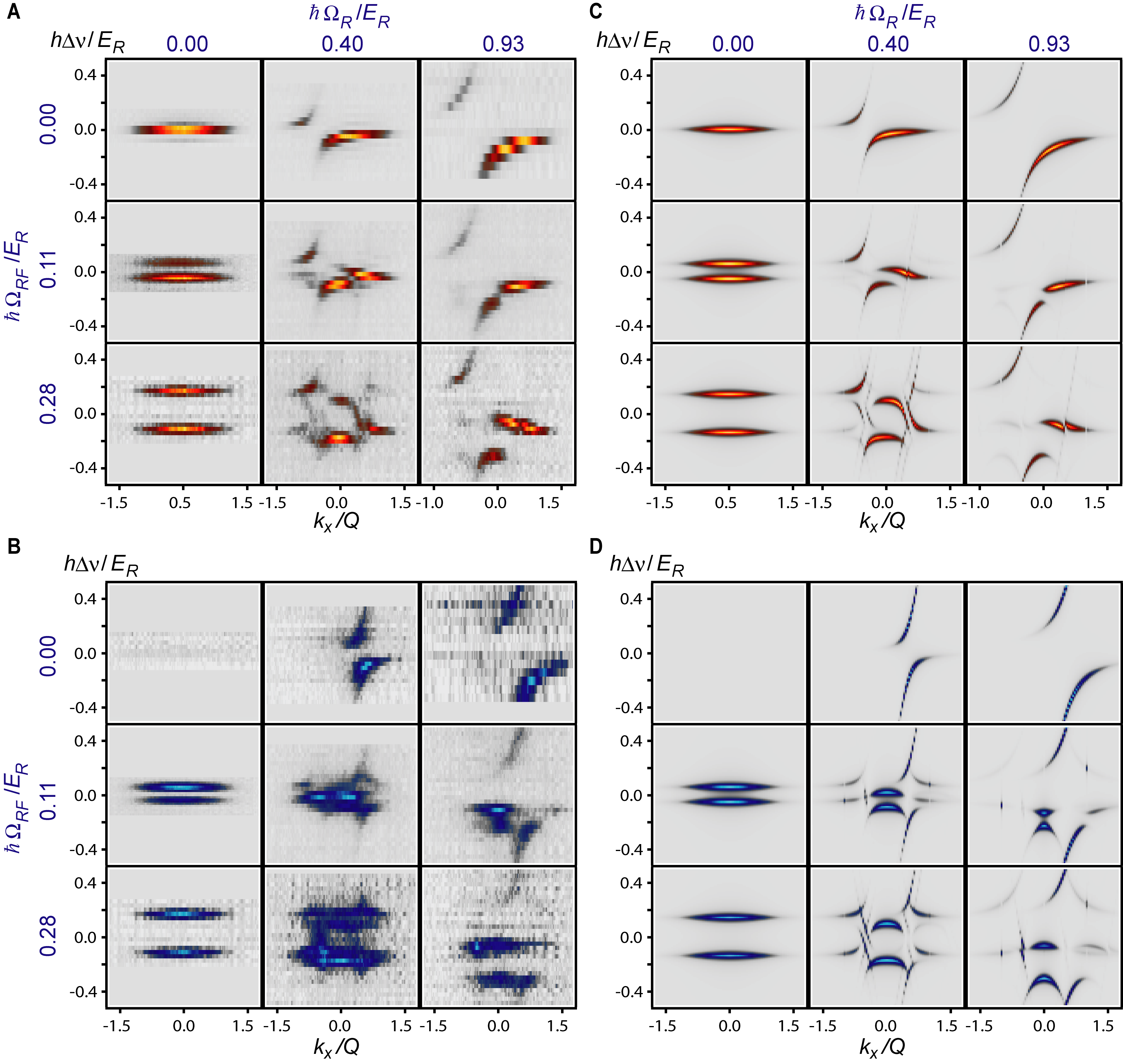}
\caption{Experimental and theoretical Spin-injection spectra of Raman/RF system for different Raman/RF strengths. (A and B) The experimental Raman/RF spin-injection spectra for injection from $\left|\downarrow\right\rangle_R$ for channels $\left|\downarrow\right\rangle$ and $\left|\uparrow \right\rangle$, respectively. The color map used is the same as Fig.~2B and 2E after rescaling to the maximum intensity (which for 4C is 20$\%$ of 4E), except for the top left panel in (B), which is scaled to the maximum intensity of the corresponding panel in (A). (C and D) Theoretical spectra corresponding to B and C. We do not take into account finite imaging resolution, which affects the sharpness along the momentum axis $k_x/Q$. Scattering due to residual interactions between hyperfine states can also lead to blurring of momenta along $k_x/Q$, and will be most pronounced for the spectra in B, as state $\left|\uparrow\right\rangle$ scatters more strongly with $\left|\downarrow\right\rangle_R$ (scattering length $\sim-450 a_0$) than $\left|\downarrow\right\rangle$ does with $\left|\downarrow\right\rangle_R$ (scattering length $\sim -2 a_0$ with typical initial $1/k_F \sim 3000 a_0$ before TOF).
}
\label{figS3f}
\end{figure*}

\subsection{Reconstructing the spinful band structure for the Raman/RF System}
We first describe a general procedure to reconstruct the band structure for any spinful lattice from spin-injection spectra. One performs spin-injection spectroscopy with reservoir states that are filled up to a Fermi momentum smaller than half the recoil momentum, $k_F<Q/2$. One then selects a prominent feature on the spectrum and finds the resonant transfer frequencies as a function of $k_x$, over one Brillouin zone. A mask centered on the resonant frequencies and repeated over all $k_x$ with period $Q$ is then created. With the mask for a specific feature applied, the spin channel $\sigma'$ for the spectrum starting from $\left|\sigma\right \rangle_R$ has transfer intensity given by
\begin{equation}
I_{\sigma',\sigma}(n,l,k_x) \propto \sum_{j} n_{\sigma} (\tilde{k}_x + l Q) |c^{(n)}_{\sigma}(\tilde{k}_x + j Q)|^2 |c^{(n)}_{\sigma'}(k_x)|^2.
\end{equation}
Here, the Fermi momentum $k_F$ is less than half the recoil momentum $Q/2$ and therefore only the $l=0$ contributes. 
Defining $\mathcal{N}_{\sigma}(n,l,\tilde{k}_x) = \sum_{j,\sigma'} I_{\sigma',\sigma}(n,l,\tilde{k}_x+jQ)$, one obtains
\begin{equation}
|c^{(n)}_{\sigma'}(k_x)|^2 = \frac{I_{\sigma',\sigma}(n,l,k_x)}{\mathcal{N}_{\sigma}(n,l,\tilde{k}_x)}.
\end{equation}
Defining $\vec{S}(k_x) = \frac{1}{2}\mathbf{c}^{(n)}(k_x)^\dag \mathbf{\sigma} \mathbf{c}^{(n)}(k_x)$, where $ \mathbf{c}^{(n)}(k_x) = (c^{(n)}_{\uparrow}(k_x), c^{(n)}_{\downarrow}(k_x))^T$, allows to measure $S_z(k_x)$. With additional RF pulses, one can measure $S_x(k_x)$ and $S_y(k_x)$. After the bands that coupled strongly to $\left|\sigma,k\right \rangle,k<Q$ are measured, one iterates the process with a larger Fermi sea to obtain other bands.

In the text, we demonstrate extraction of band structure and spin texture in a spin-orbit coupled lattice. The initial Fermi sea has $k_F>Q$, therefore spectral features corresponding to transitions out of different non-interacting reservoir bands can all appear near zero RF detuning. One can however identify spectral features corresponding to different transitions and apply the above procedure. In addition, since the Raman/RF system is set to $\delta=0$, one can invoke the additional symmetry of the band structure about $q=0$, meaning that it is sufficient to fit a certain spectral feature over half of a Brillouin zone. Due to a larger signal in the $\left|\downarrow \right\rangle$ channel, features in this channel were used to create the mask.

In order to be consistent with the earlier convention for the Rashba-Dresselhaus system, we apply a global spin rotation such that $S_y(k_x)$ corresponds to $\frac{1}{2}\left(|c^{(n)}_{\uparrow}(k_x)|^2 - |c^{(n)}_{\downarrow}(k_x)|^2\right)$. The experimentally measured $S_y(k_x)$ for the lowest band for $\hbar \Omega_R=0.93(7)E_R$ and $\hbar \Omega_{RF}=0.28(2)E_R$ is shown in Fig.~4F. Since the Hamiltonian only has components in the pseudo-spin $\hat{y}$ and $\hat{z}$ direction, one can also extract $|S_z(k_x)| = \sqrt{\frac{1}{4}-S_y(k_x)^2}$, as shown in Fig.~4E.

\end{document}